\documentclass[review, sort&compress]{elsarticle}
\usepackage{epsfig,amsmath,graphicx,amssymb,overpic}
\usepackage{bm}
\usepackage{mathrsfs}
\usepackage{graphicx}
\usepackage{epsfig}
\usepackage{dcolumn}
\usepackage{bm}
\usepackage{amsmath,amssymb,amsthm}
\usepackage[colorlinks=true,linkcolor=blue]{hyperref}
\usepackage{subfigure}
\usepackage{booktabs}
\usepackage[mathscr]{euscript}
\usepackage{ulem}
\usepackage{setspace}
\usepackage{color}
\usepackage{geometry}  
\newcommand{\bea}{\begin{eqnarray}}
\newcommand{\eea}{\end{eqnarray}}
\newcommand{\beq}{\begin{equation}}
\newcommand{\eeq}{\end{equation}}

\def\/{\over}

\usepackage{CJK}
\begin{document}

\title{Restricted $K$- and $\tau$-symmetries on multisoliton manifolds of the potential BKP and CKP equations}

\author[rvt1]{Xiazhi Hao}
\author[rvt1]{Mingxia Ji}%
\author[rvt2]{S. Y. Lou\corref{cor1}}
\cortext[cor1]{Corresponding author. Email: lousenyue@nbu.edu.cn(S. Y. Lou) }

\address[rvt1]{School of Mathematical Sciences, Zhejiang University of Technology, Hangzhou, 310014, China}
\address[rvt2]{School of Physical Science and Technology, Ningbo University, Ningbo, 315211, China}

\begin{abstract}
		We investigate the reduction of generalized $K$-symmetries and $\tau$-symmetries on the multisoliton manifolds of the potential BKP and CKP hierarchies. The  $n$-soliton solution of BKP depends on wave-number parameters $k_m$, transverse parameters $l_m$, and phase parameters $\xi_{m0}$, defining a $3n$-dimensional manifold. Its restricted $K$-symmetries project onto an $n$-dimensional phase-shift subspace, while the restricted $\tau$-symmetries engage the full $3n$-dimensional tangent space. This structure naturally yields a row-replacement determinant representation. In contrast, the  $n$-soliton solution of CKP has no independent transverse parameters, reducing the manifold dimension to $2n$. The restricted $K$-symmetries again form an $n$-dimensional phase-shift representation, but the $\tau$-symmetries now involve only $2n$ tangent directions. For CKP, the natural formulation treats elementary tangent symmetries as unknowns, producing a column-replacement determinant representation via Cramer's rule. The difference between these two determinant structures reflects not a difference in the underlying linear algebra, but rather the distinct parameterizations of the two soliton manifolds. Both formulas express explicitly how infinite-dimensional symmetry hierarchies collapse onto finite-dimensional multisoliton families.
\end{abstract}


\maketitle
\large
\doublespacing

\section{Introduction}

Integrable hierarchies of Kadomtsev–Petviashvili (KP) type admit several distinguished reductions characterized by involutive constraints on their Lax operators. The BKP and CKP hierarchies arise from orthogonal-type and symplectic-type reductions, respectively, and occupy central positions in the classification of $(2+1)$-dimensional integrable systems \cite{Date1981KP, bbib6, JimboMiwa1983,  Zabrodin2021}. Both hierarchies possess rich infinite-dimensional symmetry algebras governing their solution spaces \cite{wbib28, Tu2007, ChengHe2014}. A natural question arises: how do these infinite hierarchies of generalized symmetries act when restricted to finite-dimensional multisoliton families?

The multisoliton solutions of any integrable hierarchy form finite-dimensional families embedded in the generally infinite-dimensional solution space. Each family is parametrized by spectral and phase parameters whose number depends on the soliton count and the structure of the hierarchy. Differentiation with respect to these parameters produces natural tangent directions along the family. Since the soliton family has finite dimension, any symmetry tangent to the family must lie in the finite-dimensional space spanned by these parameter-induced directions. This observation provides a concrete framework for studying symmetry reduction: express each restricted symmetry as a linear combination of the elementary tangent symmetries, and analyze the resulting coefficient structure \cite{lou1, 25pla, mbib26, mbib28, 26zamp}.

Although the symmetry hierarchy contains infinitely many generators, their restrictions to a finite-dimensional soliton manifold must satisfy linear dependence relations. The form of these relations depends on both the parameterization of the soliton solution and the algebraic structure of the symmetry under consideration \cite{hbib9, hbib1,Orlov1986}. For the $K$-symmetries, which correspond to pure time flows in the hierarchy, the reduction typically involves only phase-shift tangent directions. For the $\tau$-symmetries, which incorporate explicit coordinate dependence, the reduction generically engages variations of all soliton parameters.

In this work we examine these reductions for the potential BKP and potential CKP equations. The BKP parameterization includes both wave-number parameters $k_m$ and independent transverse parameters $l_m$, yielding a $3n$-dimensional manifold for $n$ solitons. The CKP parameterization has no independent transverse parameter. The transverse propagation is completely determined by the wave number through the dispersion relation, reducing the manifold dimension to $2n$. This structural difference has profound consequences for the restricted symmetry algebras.

Our main objective is to derive explicit determinant formulas expressing higher-order restricted symmetries through selected bases of lower-order ones. We will show that although the symmetry reductions of both hierarchies can be characterized by determinants, their specific forms reflect the geometric structures of the respective soliton manifolds. These determinant formulas arise from linear dependence among restricted symmetries and are obtained via Cramer's rule under nondegeneracy assumptions on the coefficient matrices.

The paper is organized as follows. Section \ref{sec2} introduces the general framework of restriction to multisoliton manifolds and the construction of parameter-induced tangent symmetries. Section \ref{sec3} treats the potential BKP equation: we derive the restricted $K$- and $\tau$-symmetries, establish their closed-form coefficient formulas via recurrence relations, and construct the corresponding determinant representations. Section \ref{sec4} presents parallel results for the potential CKP equation, highlighting the simplified structure arising from the absence of transverse parameters. Section \ref{sec5} discusses the relation between the two formulations and their geometric interpretation and concludes with remarks on the broader context and future directions.

\section{Restriction to the Multisoliton Manifold and Tangent Symmetries}\label{sec2}

Let $\mathcal{M}_n$ denote the $n$-soliton manifold associated with the integrable hierarchy under consideration. Let $w_n = w_n(\mathbf{p})$ represent an $n$-soliton solution parametrized by a finite set of spectral and phase parameters $\mathbf{p} = (p_1, \dots, p_d)$, where $d$ is the dimension of the parameter space specific to the hierarchy (e.g., $d=3n$ for BKP and $d=2n$ for CKP) \cite{SatoSato1982}.

The restriction of an arbitrary generalized symmetry $F[w]$ to the soliton manifold $\mathcal{M}_n$ is defined by evaluation
\begin{equation*}
\left. F \right|_{w=w_n} \equiv F[w_n(\mathbf{p})].
\end{equation*}
Since $w_n(\mathbf{p})$ constitutes a smooth family of exact solutions, differentiation with respect to the parameters $\mathbf{p}$ yields natural tangent vectors to $\mathcal{M}_n$. Specifically, the tangent symmetry associated with the parameter $p_a$ is given by
\begin{equation*}
\sigma_{p_a} = \frac{\partial w_n}{\partial p_a}, \quad a=1, \dots, d.
\end{equation*}
By construction, each $\sigma_{p_a}$ satisfies the linearized equation on the background $w_n$. Consequently, the set $\{\sigma_{p_1}, \dots, \sigma_{p_d}\}$ spans a $d$-dimensional subspace of the tangent space $T_{w_n}\mathcal{M}_n$, which we refer to as the parameter-induced tangent space.

Consider a sequence of restricted generalized symmetries $\{F_i\}_{i \geq 1}$. Due to the finite dimensionality of $\mathcal{M}_n$, the restrictions of these symmetries must lie within the span of the elementary tangent symmetries. Thus, for each $i$, there exists a coefficient vector $\mathbf{c}_i = (c_{i,1}, \dots, c_{i,d}) \in \mathbb{R}^d$ such that
\begin{equation*}
\left. F_i \right|_{w=w_n} = \sum_{a=1}^d c_{i,a} \, \sigma_{p_a}.
\end{equation*}
This relation formalizes the symmetry reduction mechanism: although $\{F_i\}$ forms an infinite-dimensional algebra in the full solution space, its projection onto $\mathcal{M}_n$ is confined to a finite-dimensional linear subspace.

The explicit determinant representations for higher-order symmetries arise from the linear dependence relations among the coefficient vectors $\{\mathbf{c}_i\}$. The structure of the resulting determinant formula depends on the choice of unknowns in the associated linear system. We distinguish two canonical constructions:

\begin{enumerate}
	\item \textbf{Row-Replacement Formulation:} When the tangent symmetries $\{\sigma_{p_a}\}$ are treated as the basis and the symmetry coefficients as data, one considers the linear system where the rows of a $d \times d$ matrix $M$ correspond to the coefficient vectors of a selected basis of symmetries $\{F_1, \dots, F_d\}$. For any higher-order symmetry $F_k$ ($k > d$), its coefficient row $\mathbf{c}_k$ can be expressed as a linear combination of the basis rows. By Cramer's rule, the expansion coefficients are given by ratios of determinants where $\mathbf{c}_k$ replaces the corresponding rows of $M$. This form is naturally adopted for the $\tau$-symmetries of the BKP hierarchy.
	
	\item \textbf{Column-Replacement Formulation:} Alternatively, when the elementary tangent symmetries $\{\sigma_{p_a}\}$ themselves are the unknowns, one arranges the system as $\mathbf{T} = M \boldsymbol{\sigma}$, where $\mathbf{T}$ is the vector of restricted symmetries and $\boldsymbol{\sigma}$ is the vector of tangent modes. In this setting, solving for $\boldsymbol{\sigma}$ via Cramer's rule involves replacing the columns of $M$. Substituting these expressions back into the expansion of a higher-order symmetry yields a column-replacement determinant identity. This form is intrinsic to the $\tau$-symmetries of the CKP hierarchy due to its specific parameterization.
\end{enumerate}

In both cases, the validity of the determinant representation is contingent upon the non-singularity of the coefficient matrix, i.e., $\det M \neq 0$. This condition ensures that the selected subset of restricted symmetries forms a basis for the tangent space $T_{w_n}\mathcal{M}_n$. The subsequent sections will apply these two formulations to the BKP and CKP hierarchies respectively, deriving explicit finite-dimensional reductions governed by the dispersion relations of the underlying equations.

\section{Potential BKP Hierarchy: Restricted Symmetries and Determinant Representations}\label{sec3}

\subsection{The Potential BKP Equation and Its Symmetry Hierarchy}

We consider the potential BKP equation
\begin{equation}\label{BKP}
w_{xt} = 5w_{yy} - \left( w_{5x} + 15w_x w_{xxx} + 15w_x^3 - 15w_x w_y - 5w_{xxy} \right)_x.
\end{equation}
This equation constitutes the first nontrivial member of the potential BKP hierarchy, an integrable system of B-type characterized by an orthogonal reduction constraint on its pseudo-differential Lax operator. The BKP hierarchy occupies a distinguished position in the classification of $(2+1)$-dimensional integrable systems, arising from the Date-Jimbo-Kashiwara-Miwa construction via Clifford algebra methods. In the limit where solutions become independent of the transverse variable $y$, equation \eqref{BKP} reduces to the Sawada Kotera equation
$$
w_{xt} = -\left( w_{5x} + 15w_x w_{xxx} + 15w_x^3 \right)_x,
$$
a prominent $(1+1)$-dimensional integrable system with a rich structure of conservation laws and soliton solutions.

A fundamental algebraic feature distinguishing the BKP hierarchy from the standard KP hierarchy is the constraint imposed on the admissible time flows. Due to the B-type reduction, only odd-indexed flows in the Lax hierarchy are compatible with the orthogonal structure. Nevertheless, as demonstrated below, one can construct a complete sequence of generalized symmetry generators indexed by all positive integers, extending beyond the intrinsic Lax flows. This extended sequence provides the natural framework for studying symmetry reductions on multisoliton manifolds.

The generalized symmetry hierarchy $\{K_i\}_{i \geq 1}$ is generated recursively via the master-symmetry commutator relation \cite{bbib7, ybib23}
\begin{equation}\label{kms}
K_{n+1} = \frac{1}{2n-1} [\tau_4, K_n], \quad n \geq 0, \quad K_0 = \frac{1}{3},
\end{equation}
where the bracket $[\cdot, \cdot]$ denotes the commutator of evolutionary vector fields, and the master symmetry $\tau_4$ is given by
\begin{equation*}
\tau_4 = 5t K_4 + y K_3 + x K_2 + 3\left( w w_x + w_{xx} + \partial_x^{-1} w_y \right).
\end{equation*}
Here $\partial_x^{-1}$ denotes a fixed antiderivative operator with respect to $x$, defined consistently with the hierarchy's construction via the method of pseudo-differential operators. The first several members of the $K$-symmetry sequence read \cite{lou1993, 22mmas}
\begin{align*}
K_1 &= w_x, \\
K_2 &= 3w_y, \\
K_3 &= -w_{5x} + 5\partial_x^{-1} w_{yy} - 15w_x w_{3x} - 15w_x^3 + 5w_{xxy} + 15w_x w_y, \\
K_4 &= -w_{7x} + 21\bigl( w_{xyy} - w_{xxx}^2 - w_{xx} w_{4x} - w_x w_{5x} - w_x w_{xxy} - 3w_x^4 \nonumber \\
& - 3w_x w_{xx}^2 - 6w_x^2 w_{xxx} + w_x \partial_x^{-1} w_{yy} + \partial_x^{-1}(w_x w_{yy}) \nonumber \\
& + 2\partial_x^{-1}(w_{xx} w_{xxy}) - 3\partial_x^{-1}(w_x^2 w_{xy}) \bigr) + 7\partial_x^{-2} w_{yyy} + \frac{63}{2} w_y^2.
\end{align*}
Here $K_1$ generates translations in $x$, $K_2$ generates translations in $y$, and $K_3$ generates the potential BKP flow itself.
These generators form a commutative algebra under the Lie bracket of evolutionary vector fields, reflecting the integrability of the underlying system.

Parallel to the $K$-symmetry hierarchy, there exists an associated sequence of $\tau$-symmetries $\{\tau_i\}_{i \geq 1}$, which incorporate explicit polynomial dependence on the independent variables $(x, y, t)$. These symmetries play a crucial role in the algebraic structure of the hierarchy and in the construction of additional conservation laws. The initial members are
\begin{align*}
\tau_1 &= 5y K_1 + \frac{y}{3} , \\
\tau_2 &= 5t K_2 + y K_1 + \frac{x}{3} , \\
\tau_3 &= 5t K_3 + y K_2 + x K_1 + w, \\
\tau_4 &= 5t K_4 + y K_3 + x K_2 + 3\left( w w_x + w_{xx} + \partial_x^{-1} w_y \right), \\
\tau_5 &= 5t K_5 + y K_4 + x K_3 + A_5,
\end{align*}
with
$$
A_5 = -w_{4x} - 15w_x w_{xx} + 23w_{xy} + 9w w_y + 9w_x \partial_x^{-1} w_y + 5\partial_x^{-2} w_{yy} + 15\partial_x^{-1}\left( w_{xx}^2 - w_x^3 + w_x w_y \right).
$$
Higher-order $\tau$-symmetries satisfy the analogous recursion
\begin{equation}\label{taun}
\tau_{n+1} = \frac{1}{2(n-4)} [\tau_4, \tau_n], \quad n \geq 5.
\end{equation}
The $\tau$-symmetries will be shown to exhibit a richer reduction structure on the multisoliton manifold compared to the $K$-symmetries, as they involve variations with respect to all soliton parameters rather than phase shifts alone.

\subsection{Multisoliton Solutions and Parameter-Induced Tangent Symmetries}

The $n$-soliton solution of the potential BKP equation admits the logarithmic representation \cite{yao2026}
\begin{equation*}
w_n = 2(\ln f_n)_x,
\end{equation*}
where the $\tau$-function $f_n$ is constructed via the Hirota bilinear formalism. Specifically, $f_n$ takes the determinantal form
\begin{equation*}
f_n = \sum_{\mu \in \{0,1\}^n} \exp\left( \sum_{j=1}^n \mu_j \xi_j + \sum_{1 \leq j < p \leq n} \mu_j \mu_p \theta_{jp} \right).
\end{equation*}
Here the summation extends over all binary $n$-tuples $\mu = (\mu_1, \dots, \mu_n)$, encoding the presence ($\mu_j = 1$) or absence ($\mu_j = 0$) of each soliton component. This exponential representation is characteristic of the direct scattering approach to soliton theory and provides an explicit algebraic description of the multisoliton manifold.

The individual soliton phase functions are given by
\begin{equation*}
\xi_j = k_j x + l_j y - \frac{k_j^6 - 5k_j^3 l_j - 5l_j^2}{k_j} t + \xi_{j0}, \quad j=1,\dots,n,
\end{equation*}
where $(k_j, l_j)$ are the spectral parameters encoding the wave number and transverse wave number of the $j$-th soliton, respectively, and $\xi_{j0}$ represents an arbitrary phase constant. The dispersion relation embedded in this formula,
$$
\omega_j = -\frac{k_j^6 - 5k_j^3 l_j - 5l_j^2}{k_j}
$$
characterizes the fifth-order dispersion structure of the BKP equation. The phase-shift coefficients encoding pairwise soliton interactions are specified by
\begin{normalsize}\begin{equation*}\begin{split}
e^{\theta_{jp}} = \frac{k_j k_p (k_j - k_p)\bigl[k_j k_p (k_j - k_p)(k_j^2 - k_j k_p + k_p^2) - k_j(k_j l_p + 2k_p l_j) + k_p(2k_j l_p + k_p l_j)\bigr] + (k_j l_p - k_p l_j)^2}{k_j k_p (k_j + k_p)\bigl[k_j k_p (k_j + k_p)(k_j^2 + k_j k_p + k_p^2) - k_j(k_j l_p + 2k_p l_j) - k_p(2k_j l_p + k_p l_j)\bigr] + (k_j l_p - k_p l_j)^2}.
\end{split}
\end{equation*}\end{normalsize}
These interaction coefficients are determined uniquely by requiring that $w_n$ satisfy the BKP equation, and they encode the nontrivial phase shifts induced by soliton collisions. The parameters $(k_j, l_j, \xi_{j0})$ are constrained to a domain where all denominators  remain nonzero, ensuring regularity of the solution.

The $n$-soliton manifold $\mathcal{M}_n^{\text{BKP}}$ is therefore parametrized by the $3n$-dimensional vector
$$
\mathbf{p} = (k_1, l_1, \xi_{10}, \dots, k_n, l_n, \xi_{n0}) \in \mathbb{R}^{3n}.
$$
This parameterization naturally partitions the tangent space into three families of elementary symmetries, corresponding to infinitesimal variations of the spectral parameters, transverse parameters, and phase constants, respectively.

Differentiation with respect to these parameters yields three families of elementary tangent symmetries. The phase-shift tangent symmetries are given by
\begin{equation*}
\sigma_{\xi_{m0}} = \frac{\partial w_n}{\partial \xi_{m0}} = 2\left( \frac{f_{n,\xi_{m0}}}{f_n} \right)_x, \quad m=1,\dots,n,
\end{equation*}
where
$$
f_{n,\xi_{m0}} = \sum_{\mu \in \{0,1\}^n} \mu_m \exp\left( \sum_{j=1}^n \mu_j \xi_j + \sum_{1 \leq j < p \leq n} \mu_j \mu_p \theta_{jp} \right).
$$
This expression represents the partial derivative of the $\tau$-function with respect to $\xi_{m0}$, with the factor $\mu_m$ selecting only those terms in the sum where the $m$-th soliton is present. Geometrically, $\sigma_{\xi_{m0}}$ corresponds to an infinitesimal translation of the $m$-th soliton along its characteristic direction, leaving all other solitons and their mutual interactions unchanged.

Similarly, variations of the spectral parameters produce the tangent symmetries
\begin{equation*}
\sigma_{k_m} = \frac{\partial w_n}{\partial k_m}, \qquad \sigma_{l_m} = \frac{\partial w_n}{\partial l_m}, \quad m=1,\dots,n.
\end{equation*}
Unlike the phase-shift symmetries, these involve modifications to the soliton velocities and transverse propagation angles, as well as implicit changes in the interaction coefficients $\theta_{jp}$ whenever $j=m$ or $p=m$. The collection
$$
\mathcal{B}_n = \{\sigma_{k_1}, \sigma_{l_1}, \sigma_{\xi_{10}}, \dots, \sigma_{k_n}, \sigma_{l_n}, \sigma_{\xi_{n0}}\}
$$
forms a basis for the tangent space $T_{w_n} \mathcal{M}_n^{\text{BKP}}$, providing the natural coordinates for expressing restricted symmetries.

\subsection{Restriction of the BKP $K$-Symmetries}

A fundamental observation is that when the $K$-symmetry generators are restricted to the $n$-soliton manifold, they collapse to the lower-dimensional subspace spanned by the phase-shift tangent symmetries alone. Specifically, we have
\begin{equation}\label{Ki}
\left. K_i \right|_{w=w_n} = \sum_{m=1}^n \kappa_{i,m} \sigma_{\xi_{m0}}, \quad i \geq 1,
\end{equation}
where the coefficient functions $\kappa_{i,m}$ depend solely on the spectral parameters $(k_m, l_m)$ and are independent of the phase constants $\xi_{j0}$. This reduction reflects the fact that the $K$-symmetries correspond to time translations in the hierarchy, which act on solitons primarily by shifting their phases.

Explicit computation of the restricted symmetries for small orders yields the coefficient functions
\begin{align*}
\kappa_{1,m} &= k_m, \\[6pt]
\kappa_{2,m} &= 3l_m, \\[6pt]
\kappa_{3,m} &= -\frac{k_m^6 - 5k_m^3 l_m - 5l_m^2}{k_m}, \\[6pt]
\kappa_{4,m} &= -\frac{k_m^{9} - 21k_m^3 l_m^2 - 7l_m^3}{k_m^2}, \\[6pt]
\kappa_{5,m} &= -\frac{9l_m(k_m^9 - 3k_m^6 l_m - 6k_m^3 l_m^2 - l_m^3)}{k_m^3}, \\[6pt]
\kappa_{6,m} &= \frac{k_m^{15} - 11k_m^{12} l_m - 11k_m^{9} l_m^2 + 143k_m^{6} l_m^3 + 110k_m^3 l_m^4 + 11l_m^5}{k_m^4}.
\end{align*}
These expressions reveal a nontrivial algebraic structure involving both polynomial and rational dependence on the spectral parameters, arising from the fifth-order dispersion of the BKP equation and the complexity of the interaction coefficients.

A key structural property is that these coefficients satisfy the three-term linear recurrence relation
\begin{equation*}
\kappa_{i+2,m} = \frac{k_m^3 + 2l_m}{k_m} \kappa_{i+1,m} - \frac{(k_m^3 - l_m)^2}{k_m^2} \kappa_{i,m}, \quad i \geq 1,
\end{equation*}
with initial conditions $\kappa_{1,m} = k_m$ and $\kappa_{2,m} = 3l_m$. This recurrence arises from the recursive structure of the master-symmetry commutator \eqref{kms} and provides a computational framework for determining all higher-order coefficients. The characteristic equation associated with this recurrence is
$$
\lambda^2 - \frac{k_m^3 + 2l_m}{k_m} \lambda + \frac{(k_m^3 - l_m)^2}{k_m^2} = 0,
$$
whose roots are
\begin{equation}\label{cr}
\lambda_{\pm,m} = \frac{k_m^3 + 2l_m \pm \sqrt{3k_m^3(4l_m - k_m^3)}}{2k_m}.
\end{equation}
The discriminant $\Delta_m = 3k_m^3(4l_m - k_m^3)$ determines the nature of the solutions: when $\Delta_m > 0$ the roots are real and distinct, when $\Delta_m = 0$ they coincide, and when $\Delta_m < 0$ they form a complex conjugate pair.

For distinct roots, the general solution of the recurrence is obtained by standard methods. Writing $\kappa_{i,m} = a_m \lambda_{+,m}^{i-1} + b_m \lambda_{-,m}^{i-1}$ and imposing the initial conditions yields the system
$$
a_m + b_m = k_m, \qquad a_m \lambda_{+,m} + b_m \lambda_{-,m} = 3l_m.
$$
Solving for $a_m$ and $b_m$ produces the closed-form expression
\begin{equation}\label{kim}
\kappa_{i,m} = \frac{(3l_m - k_m \lambda_{-,m}) \lambda_{+,m}^{i-1} + (k_m \lambda_{+,m} - 3l_m) \lambda_{-,m}^{i-1}}{\lambda_{+,m} - \lambda_{-,m}}, \quad i \geq 1.
\end{equation}
When the discriminant vanishes (i.e., $4l_m = k_m^3$), the roots coalesce to $\lambda_m = 3k_m^2/4$, and the solution reduces to the repeated-root form
$$
\kappa_{i,m} = k_m \lambda_m^{i-1} + (i-1)(3l_m - k_m \lambda_m) \lambda_m^{i-2}, \quad i \geq 1.
$$
However, substituting $\lambda_m = 3k_m^2/4$ and $l_m = k_m^3/4$ into this formula shows that the second term vanishes, yielding simply
$$
\kappa_{i,m} = k_m \lambda_m^{i-1}, \quad i \geq 1.
$$
Thus the apparent singularity in \eqref{kim} at coincident roots is removable, and the closed form remains valid in the limit $\lambda_{+,m} \to \lambda_{-,m}$.

Consequently, the restricted $K$-symmetries admit the unified closed-form representation
\begin{equation*}
\left. K_i \right|_{w=w_n} = \sum_{m=1}^n \frac{(3l_m - k_m \lambda_{-,m}) \lambda_{+,m}^{i-1} + (k_m \lambda_{+,m} - 3l_m) \lambda_{-,m}^{i-1}}{\lambda_{+,m} - \lambda_{-,m}} \sigma_{\xi_{m0}}, \quad i \geq 1,
\end{equation*}
valid for arbitrary positive integer $i$. This formula demonstrates explicitly how the infinite-dimensional $K$-symmetry algebra projects onto the finite-dimensional phase-shift tangent space, with coefficients determined entirely by the dispersion relation encoded in the characteristic roots \eqref{cr}.


The linear dependence structure of the restricted $K$-symmetries can be made explicit through determinant formulas, which express higher-order symmetries as linear combinations of a selected basis. To formulate this systematically, we define the phase-tangent vector
$$
\boldsymbol{\sigma}_{\xi} = (\sigma_{\xi_{10}}, \dots, \sigma_{\xi_{n0}})^T \in \mathbb{R}^n
$$
and the vector of the first $n$ restricted $K$-symmetries
$$
\mathbf{K}_n = \left( \left. K_1 \right|_{w=w_n}, \dots, \left. K_n \right|_{w=w_n} \right)^T \in \mathbb{R}^n.
$$
The relation \eqref{Ki} can be recast as the linear system
\begin{equation}\label{ls}
\mathbf{K}_n = M_n \boldsymbol{\sigma}_{\xi},
\end{equation}
where the $n \times n$ coefficient matrix $M_n$ has entries
$$
(M_n)_{ij} = \kappa_{i,j}, \quad i,j = 1, \dots, n.
$$
Explicitly, the first few rows of $M_n$ are
$$
M_n = \begin{pmatrix}
k_1 & k_2 & \cdots & k_n \\[4pt]
3l_1 & 3l_2 & \cdots & 3l_n \\[4pt]
-k_1^5 + 5k_1^2 l_1 + \dfrac{5l_1^2}{k_1} & -k_2^5 + 5k_2^2 l_2 + \dfrac{5l_2^2}{k_2} & \cdots & -k_n^5 + 5k_n^2 l_n + \dfrac{5l_n^2}{k_n} \\[6pt]
\vdots & \vdots & \ddots & \vdots
\end{pmatrix}.
$$
This matrix encodes the spectral multipliers arising from the BKP dispersion relation, and its determinant controls the linear independence of the first $n$ restricted symmetries.

Let $\Delta_n = \det M_n$. The nondegeneracy condition $\Delta_n \neq 0$ is generically satisfied in the parameter space and ensures that the tangent symmetries $\{\sigma_{\xi_{10}}, \dots, \sigma_{\xi_{n0}}\}$ can be uniquely recovered from the restricted symmetries $\{K_1|_{w=w_n}, \dots, K_n|_{w=w_n}\}$. Geometrically, this condition states that the first $n$ restricted $K$-symmetries provide a linearly independent spanning set for the $n$-dimensional phase-shift subspace.

For $m=1,\dots,n$, denote by $M_{n,m}$ the matrix obtained by replacing the $m$-th column of $M_n$ with the vector $\mathbf{K}_n$, and define
$$
\Delta_{n,m} = \det M_{n,m}.
$$
By Cramer's rule applied to the linear system \eqref{ls}, we obtain
\begin{equation*}
\sigma_{\xi_{m0}} = \frac{\Delta_{n,m}}{\Delta_n}, \quad m=1,\dots,n,
\end{equation*}
expressing each elementary tangent symmetry as a ratio of determinants. Substituting this into the expansion \eqref{Ki} for a higher-order symmetry $K_i$ with $i \geq n+1$ yields the determinant representation
\begin{equation}\label{DK}
\left. K_i \right|_{w=w_n} = \frac{1}{\Delta_n} \sum_{m=1}^n \kappa_{i,m} \Delta_{n,m}, \quad i \geq n+1.
\end{equation}
Combining with the closed-form expression \eqref{kim} for $\kappa_{i,m}$, we arrive at the explicit formula
\begin{equation*}
\left. K_i \right|_{w=w_n} = \frac{1}{\Delta_n} \sum_{m=1}^n \frac{(3l_m - k_m \lambda_{-,m}) \lambda_{+,m}^{i-1} + (k_m \lambda_{+,m} - 3l_m) \lambda_{-,m}^{i-1}}{\lambda_{+,m} - \lambda_{-,m}} \Delta_{n,m}, \quad i \geq n+1.
\end{equation*}

This formula constitutes the main result for the restricted $K$-symmetries of the BKP hierarchy. It demonstrates that every higher-order restricted $K$-symmetry is uniquely determined by the first $n$ restricted symmetries through a determinant identity, with coefficients governed by the characteristic roots of the dispersion relation. The structure of this representation is characteristic of finite-dimensional reductions of infinite-dimensional symmetry algebras on soliton manifolds: the algebraic complexity of the coefficients encodes the dispersion properties of the underlying integrable system, while the determinant structure reflects the generic linear independence of the elementary tangent modes.

\subsection{Restriction of the BKP $\tau$-Symmetries}

In contrast to the $K$-symmetries, which restrict to the phase-shift subspace, the $\tau$-symmetries involve the full $3n$-dimensional tangent space of the soliton manifold. This richer structure arises because the $\tau$-symmetries incorporate explicit coordinate dependence and consequently couple to variations of all soliton parameters, not merely phase shifts.

For each $i \geq 3$, the restriction of $\tau_i$ to the $n$-soliton manifold takes the form
\begin{equation}\label{taui}
\left. \tau_i \right|_{w=w_n} = \sum_{m=1}^n \left( A_{i,m} \sigma_{k_m} + B_{i,m} \sigma_{l_m} + C_{i,m} \sigma_{\xi_{m0}} \right),
\end{equation}
where the coefficients $(A_{i,m}, B_{i,m}, C_{i,m})$ encode the projections onto the three tangent directions associated with the $m$-th soliton. These coefficients are functions of the spectral parameters $(k_m, l_m)$ and the symmetry order $i$, but remain independent of the phase constants $\xi_{j0}$.

Explicit evaluation of the first several restricted $\tau$-symmetries yields
\begin{align}
\left. \tau_3 \right|_{w=w_n} &= \sum_{m=1}^n \left( k_m \sigma_{k_m} + 3l_m \sigma_{l_m} \right), \\[4pt]\label{tau4}
\left. \tau_4 \right|_{w=w_n} &= \sum_{m=1}^n \left( 3l_m \sigma_{k_m} - \frac{k_m^6 - 5k_m^3 l_m - 5l_m^2}{k_m} \sigma_{l_m} + 3k_m^2 \sigma_{\xi_{m0}} \right), \\[4pt]\label{tau5}
\left. \tau_5 \right|_{w=w_n} &= \sum_{m=1}^n \left( -\frac{k_m^6 - 5k_m^3 l_m - 5l_m^2}{k_m} \sigma_{k_m} - \frac{k_m^9 - 21k_m^3 l_m^2 - 7l_m^3}{k_m^2} \sigma_{l_m} + 18k_m l_m \sigma_{\xi_{m0}} \right).
\end{align}
These expressions reveal several structural features. First, the coefficient of $\sigma_{k_m}$ in $\tau_{i+1}$ equals the coefficient of $\sigma_{l_m}$ in $\tau_i$, reflecting a shift property inherited from the recursion \eqref{taun}. Second, unlike the $K$-symmetries, the $\tau$-symmetries generically have nonzero components in all three tangent directions, indicating that they generate infinitesimal deformations of the entire soliton parameter space.

The coefficients satisfy coupled recurrence relations that mirror the structure of the $K$-symmetry recursion. Specifically, we have
\begin{equation}\label{Ai}
A_{i+2,m} = \frac{k_m^3 + 2l_m}{k_m} A_{i+1,m} - \frac{(k_m^3 - l_m)^2}{k_m^2} A_{i,m}, \quad B_{i,m} = A_{i+1,m}, \quad i \geq 3,
\end{equation}
with initial conditions $A_{3,m} = k_m$ and $A_{4,m} = 3l_m$. The relation $B_{i,m} = A_{i+1,m}$ expresses the aforementioned shift property. Solving the recurrence for $A_{i,m}$ yields the closed form
\begin{equation}\label{Aii}
A_{i,m} = \frac{(3l_m - k_m \lambda_{-,m}) \lambda_{+,m}^{i-3} + (k_m \lambda_{+,m} - 3l_m) \lambda_{-,m}^{i-3}}{\lambda_{+,m} - \lambda_{-,m}}, \quad i \geq 3,
\end{equation}
where $\lambda_{\pm,m}$ are the characteristic roots defined in \eqref{cr}. Comparing with \eqref{kim}, we observe that $A_{i,m} = \kappa_{i-2,m}$, establishing a direct connection between the $\tau$-symmetry coefficients and the $K$-symmetry coefficients.

The coefficient $C_{i,m}$ encoding the phase-shift component has a more intricate structure. It is given by
\begin{equation}\label{Ci}
C_{i,m} = (i-3)\left(\lambda_{+,m}^{i-3} + \lambda_{-,m}^{i-3}\right) + 2\frac{\lambda_{+,m}^{i-2} - \lambda_{-,m}^{i-2}}{\lambda_{+,m} - \lambda_{-,m}} - \frac{2}{k_m} A_{i,m}, \quad i \geq 3.
\end{equation}
This expression reflects the explicit dependence of the $\tau$-symmetries on the hierarchy order $i$, arising from their coordinate-dependent structure. The first two terms involve symmetric and antisymmetric combinations of the characteristic roots, while the final term couples back to the coefficient $A_{i,m}$.

To construct the determinant representation for the $\tau$-symmetries, we introduce the $3n$-dimensional vector of elementary tangent symmetries
$$
\boldsymbol{\sigma} = (\sigma_{k_1}, \sigma_{l_1}, \sigma_{\xi_{10}}, \sigma_{k_2}, \sigma_{l_2}, \sigma_{\xi_{20}}, \dots, \sigma_{k_n}, \sigma_{l_n}, \sigma_{\xi_{n0}})^T,
$$
and define the coefficient row vector for the $i$-th $\tau$-symmetry as
$$
\mathbf{v}_i = (A_{i,1}, B_{i,1}, C_{i,1}, A_{i,2}, B_{i,2}, C_{i,2}, \dots, A_{i,n}, B_{i,n}, C_{i,n}) \in \mathbb{R}^{3n}.
$$
The restriction formula \eqref{taui} is then equivalently written as
$$
\left. \tau_i \right|_{w=w_n} = \mathbf{v}_i \cdot \boldsymbol{\sigma}.
$$

We select the first $3n$ restricted $\tau$-symmetries, corresponding to $i = 3, 4, \dots, 3n+2$, and form the $3n \times 3n$ coefficient matrix
$$
M_n^{\tau} = \begin{pmatrix}
\mathbf{v}_3 \\
\mathbf{v}_4 \\
\vdots \\
\mathbf{v}_{3n+2}
\end{pmatrix}.
$$
Explicitly, this matrix has the block structure
$$
M_n^{\tau} = \begin{pmatrix}
A_{3,1} & B_{3,1} & C_{3,1} & A_{3,2} & B_{3,2} & C_{3,2} & \cdots & A_{3,n} & B_{3,n} & C_{3,n} \\[4pt]
A_{4,1} & B_{4,1} & C_{4,1} & A_{4,2} & B_{4,2} & C_{4,2} & \cdots & A_{4,n} & B_{4,n} & C_{4,n} \\[4pt]
\vdots & \vdots & \vdots & \vdots & \vdots & \vdots & \ddots & \vdots & \vdots & \vdots \\[4pt]
A_{3n+2,1} & B_{3n+2,1} & C_{3n+2,1} & A_{3n+2,2} & B_{3n+2,2} & C_{3n+2,2} & \cdots & A_{3n+2,n} & B_{3n+2,n} & C_{3n+2,n}
\end{pmatrix}.
$$
Each row of this matrix encodes the tangent-space decomposition of one restricted $\tau$-symmetry, while each block of three columns corresponds to the three tangent directions associated with a single soliton.

Let $\Delta_n^{\tau} = \det M_n^{\tau}$, and assume $\Delta_n^{\tau} \neq 0$. This nondegeneracy condition ensures that the selected $3n$ restricted $\tau$-symmetries form a basis for the full tangent space $T_{w_n} \mathcal{M}_n^{\text{BKP}}$. Under this assumption, any higher-order restricted $\tau$-symmetry can be expressed as a linear combination of the basis elements.

For any $i \geq 3n+3$, the coefficient row $\mathbf{v}_i$ can be expanded in terms of the rows of $M_n^{\tau}$. Denote by $M_{n,r}^{\tau,(i)}$ the matrix obtained by replacing the $r$-th row of $M_n^{\tau}$ with $\mathbf{v}_i$, and define
$$
\Delta_{n,r}^{\tau,(i)} = \det M_{n,r}^{\tau,(i)}, \quad r = 1, \dots, 3n.
$$
By the row-replacement form of Cramer's rule, the expansion coefficients are given by the determinant ratios $\Delta_{n,r}^{\tau,(i)} / \Delta_n^{\tau}$. Consequently, the higher-order restricted $\tau$-symmetry admits the determinant representation
\begin{equation}\label{DT}
\left. \tau_i \right|_{w=w_n} = \sum_{r=1}^{3n} \frac{\Delta_{n,r}^{\tau,(i)}}{\Delta_n^{\tau}} \left. \tau_{r+2} \right|_{w=w_n}, \quad i \geq 3n+3.
\end{equation}

This formula is the analogue for $\tau$-symmetries of the determinant representation \eqref{DK} derived for $K$-symmetries. However, the $\tau$-symmetry representation involves a $3n \times 3n$ matrix rather than an $n \times n$ matrix, reflecting the full dimensionality of the soliton parameter space. The row-replacement structure arises naturally from treating the restricted symmetries as the data vector and the elementary tangent symmetries as the unknowns to be eliminated.

Formula \eqref{DT} demonstrates that on the finite-dimensional $n$-soliton manifold, every higher-order $\tau$-symmetry is uniquely determined by the first $3n$ restricted $\tau$-symmetries, with expansion coefficients given by determinant ratios depending on the soliton parameters through the functions $A_{i,m}$, $B_{i,m}$, and $C_{i,m}$. This constitutes a complete description of the $\tau$-symmetry reduction structure for the BKP hierarchy, subject to the nondegeneracy condition $\Delta_n^{\tau} \neq 0$.

\subsection{Illustrative Example: The One-Soliton Case}

To illustrate the mechanics of the determinant representation concretely, we consider the simplest nontrivial case $n=1$. For a single-soliton solution, the tangent space is three-dimensional, spanned by $\{\sigma_k, \sigma_l, \sigma_{\xi_0}\}$. The first three restricted $\tau$-symmetries yield the linear system
$$
\begin{pmatrix}
\tau_3 \\[4pt]
\tau_4 \\[4pt]
\tau_5
\end{pmatrix} 
= M_1^{\tau} 
\begin{pmatrix}
\sigma_k \\[4pt]
\sigma_l \\[4pt]
\sigma_{\xi_0}
\end{pmatrix}, 
\qquad 
M_1^{\tau} = 
\begin{pmatrix}
k & 3l & 0 \\[4pt]
3l & q & 3k^2 \\[4pt]
q & r & 18kl
\end{pmatrix},
$$
where for notational brevity we have suppressed the subscript $m=1$ and defined
$$
q = \frac{-k^6 + 5k^3 l + 5l^2}{k}, \qquad r = \frac{-k^9 + 21k^3 l^2 + 7l^3}{k^2}.
$$
These quantities represent the coefficients appearing in $\tau_4$ and $\tau_5$ according to formulas \eqref{tau4}–\eqref{tau5}.

The determinant of $M_1^{\tau}$ is computed as
$$
\Delta_1^{\tau} = \det M_1^{\tau} = 27k^2 l q - 3k^3 r - 162kl^3.
$$
Substituting the expressions for $q$ and $r$ and simplifying yields
$$
\Delta_1^{\tau} = 3k(k^3-l)(k^3-4l)^2.
$$
This is generically nonzero, confirming that the three restricted $\tau$-symmetries are linearly independent for generic parameter values $(k, l)$.

The next restricted symmetry $\tau_6$ has coefficient row $\mathbf{v}_6 = (r_6, s_6, c_6)$, where
$$
r_6 = A_{6,1}, \quad s_6 = B_{6,1} = A_{7,1}, \quad c_6 = C_{6,1}.
$$
Using the recurrence relations \eqref{Ai}–\eqref{Ci}, we compute
$$
s_6 = -\frac{9l(k^9 - 3k^6 l - 6k^3 l^2 - l^3)}{k^3}, \qquad c_6 = -3(2k^6 - 10k^3 l - 19l^2).
$$

Assuming $\Delta_1^{\tau} \neq 0$, the row-replacement formula \eqref{DT}  for $i=6$ gives
\begin{equation}\label{T6}
\tau_6 = \frac{\Delta_{1,1}^{\tau,(6)}}{\Delta_1^{\tau}} \tau_3 + \frac{\Delta_{1,2}^{\tau,(6)}}{\Delta_1^{\tau}} \tau_4 + \frac{\Delta_{1,3}^{\tau,(6)}}{\Delta_1^{\tau}} \tau_5,
\end{equation}
where
$$
\Delta_{1,1}^{\tau,(6)} = \det 
\begin{pmatrix}
r_6 & s_6 & c_6 \\[4pt]
3l & q & 3k^2 \\[4pt]
q & r & 18kl
\end{pmatrix}, 
\quad
\Delta_{1,2}^{\tau,(6)} = \det 
\begin{pmatrix}
k & 3l & 0 \\[4pt]
r_6 & s_6 & c_6 \\[4pt]
q & r & 18kl
\end{pmatrix}, 
\quad
\Delta_{1,3}^{\tau,(6)} = \det 
\begin{pmatrix}
k & 3l & 0 \\[4pt]
3l & q & 3k^2 \\[4pt]
r_6 & s_6 & c_6
\end{pmatrix}.
$$

Expanding these $3 \times 3$ determinants explicitly yields rational expressions in $(k, l)$, which upon substitution into \eqref{T6} produce the linear combination expressing $\tau_6$ in terms of $\{\tau_3, \tau_4, \tau_5\}$. This explicit example demonstrates the mechanics of the row-replacement determinant representation in the simplest nontrivial case, illustrating how the $3n$-dimensional structure manifests at $n=1$. For higher values of $n$, the same principle applies, though the computational complexity increases with the dimension $3n$ of the coefficient matrix.

\section{Potential CKP Hierarchy: Restricted Symmetries and Determinant Representations}\label{sec4}

\subsection{The Potential CKP Equation and Its Symmetry Hierarchy}

We now examine the potential CKP  equation, a distinguished member of the $C$-type integrable hierarchies arising from symplectic reduction constraints. The potential CKP equation is given by
\begin{equation}
w_{xt} = 5w_{yy} - \left( w_{5x} + 15w_x w_{xxx} + 15w_x^3 + \frac{45}{4}w_{xx}^2 - 15w_x w_y - 5w_{xxy} \right)_x.
\label{eq:pckp}
\end{equation}
This equation constitutes the first nontrivial member of the potential CKP hierarchy, characterized by a symplectic reduction that fundamentally distinguishes it from the orthogonal-type BKP hierarchy. The structural signature of this distinction is immediately visible in the presence of the $w_{xx}^2$ term, which does not appear in the corresponding BKP equation. In the limit where solutions become independent of the transverse variable $y$, equation \eqref{eq:pckp} reduces to the Kaup--Kupershmidt equation \cite{abib37}, a prominent $(1+1)$-dimensional integrable system whose dispersion structure differs fundamentally from the Sawada--Kotera equation obtained from the BKP reduction.

The symplectic constraint inherited by the CKP hierarchy restricts the admissible isospectral flows in its standard Lax formulation to odd orders only. Nevertheless, as in the BKP case, one can construct a complete sequence of generalized symmetry generators indexed by all positive integers. We begin with the generalized $K$-symmetry hierarchy $\{K_i\}_{i\geq 1}$, which consists of evolutionary vector fields commuting with the potential CKP flow. The initial members are
\begin{align*}
K_1 &= w_x, \\
K_2 &= 3w_y, \\
K_3 &= 5\partial_x^{-1} w_{yy} - \left( w_{5x} + 15w_x w_{xxx} + 15w_x^3 + \frac{45}{4}w_{xx}^2 - 15w_x w_y - 5w_{xxy} \right), \\
K_4 &= -w_{7x} + 7(\partial_x^{-2} w_{yyy} + 3\partial_x^{-1} w_x w_{yy} + 3w_x \partial_x^{-1} w_{yy} - 9\partial_x^{-1} w_x^2 w_{xy}) \\
&\quad + 21(w_{xyy} - w_x w_{xxy} - w_{xx} w_{xy} - w_x w_{5x} - 3w_x^4 - 6w_x^2 w_{xxx}) \\
&\quad + \frac{21}{4}(6w_y^2 - 7w_{xxx}^2 - 10w_{xx} w_{4x} - 30w_x w_{xx}^2 - 2\partial_x^{-1} w_{xy} w_{xxx}).
\end{align*}

Parallel to the $K$-symmetries, the CKP hierarchy admits a sequence of $\tau$-symmetries $\{\tau_i\}_{i\geq 1}$ incorporating explicit polynomial dependence on the independent variables. These are constructed systematically to satisfy certain scaling relations with the $K$-symmetries. The initial $\tau$-symmetries read
\begin{align*}
\tau_1 &= 15t K_1 + y, \notag \\
\tau_2 &= 15t K_2 + 3y K_1 + x, \notag \\
\tau_3 &= 15t K_3 + 3y K_2 + 3x K_1 + 3w, \\
\tau_4 &= 15t K_4 + 3y K_3 + 3x K_2 + 9\left(w w_x + \frac{1}{2}w_{xx} + \partial_x^{-1} w_y\right), \notag \\
\tau_5 &= 15t K_5 + 3y K_4 + 3x K_3 + A_5, \notag
\end{align*}
with
\begin{equation*}
A_5 = -\frac{15}{2}w_{4x} - 99w_x w_{xx} + 60w_{xy} + 27w w_y + 27w_x \partial_x^{-1} w_y + 15\partial_x^{-2} w_{yy} + 45\partial_x^{-1}\left(\frac{w_{xx}^2}{4} - w_x^3 + w_x w_y\right).
\end{equation*}

The master-symmetry structure of the CKP hierarchy is encoded in the observation that $\tau_4$ acts as a generating operator for both hierarchies. Specifically, the higher-order $K$-symmetries satisfy the recursion relation
\begin{equation*}
K_{n+1} = \frac{1}{6n-3} [\tau_4, K_n], \quad n \geq 0, \quad K_0 = \frac{1}{3}.
\end{equation*}
The scaling factor $6n-3$ reflects the symplectic structure and differs from the corresponding factor $2n-1$ in the BKP recursion, encoding the distinct algebraic constraints of the two hierarchies. Similarly, higher-order $\tau$-symmetries are generated via
\begin{equation*}
\tau_{n+1} = \frac{1}{6(n-4)} [\tau_4, \tau_n], \quad n \geq 5.
\end{equation*}

\subsection{Multisoliton Solutions and Parameter-Induced Tangent Symmetries}

The $n$-soliton solution of the potential CKP equation admits the logarithmic representation
\begin{equation*}
w_n = (\ln f_n)_x,
\end{equation*}
where the $\tau$-function $f_n$ takes the determinantal form
\begin{equation*}
f_n = \sum_{\mu \in \{0,1\}^n} \exp\left( \sum_{j=1}^n \mu_j \xi_j + \sum_{1 \leq j < p \leq n} \mu_j \mu_p \theta_{jp} \right).
\end{equation*}
The individual soliton phase functions are given by
\begin{equation}
\xi_j = k_j x + \frac{1}{4}k_j^3 y + \frac{9}{16}k_j^5 t + \xi_{j0}, \quad j = 1, \ldots, n.
\label{eq:ckp_phase}
\end{equation}
 The phase-shift coefficients encoding pairwise soliton interactions are specified by
\begin{equation*}
e^{\theta_{jp}} = \frac{(k_j - k_p)^2}{(k_j + k_p)^2}.
\end{equation*}

A fundamental structural distinction from the BKP case now emerges: the CKP multisoliton solution \eqref{eq:ckp_phase} contains no independent transverse parameter $l_j$ analogous to those appearing in the BKP parameterization. The transverse propagation is completely determined by the wave number $k_j$ through the coefficient $k_j^3/4$ in the $y$-dependence of $\xi_j$. This dimensional reduction has profound consequences for the restricted symmetry structure.

The $n$-soliton manifold $M_n^{\text{CKP}}$ is therefore parametrized by the $2n$-dimensional vector
\begin{equation*}
p = (k_1, \xi_{10}, k_2, \xi_{20}, \ldots, k_n, \xi_{n0}) \in \mathbb{R}^{2n}.
\end{equation*}
Differentiation with respect to these parameters yields two families of elementary tangent symmetries. The phase-shift tangent symmetries are given by
\begin{equation*}
\sigma_{\xi_{m0}} = \frac{\partial w_n}{\partial \xi_{m0}} = \left( \frac{f_{n,\xi_{m0}}}{f_n} \right)_x, \quad m = 1, \ldots, n,
\end{equation*}
where
\begin{equation*}
f_{n,\xi_{m0}} = \sum_{\mu \in \{0,1\}^n} \mu_m \exp\left( \sum_{j=1}^n \mu_j \xi_j + \sum_{1 \leq j < p \leq n} \mu_j \mu_p \theta_{jp} \right).
\end{equation*}
Similarly, variations of the spectral parameters produce the tangent symmetries
\begin{equation*}
\sigma_{k_m} = \frac{\partial w_n}{\partial k_m}, \quad m = 1, \ldots, n.
\end{equation*}
Unlike the phase-shift symmetries, these involve modifications to both the soliton velocities and the interaction coefficients $\theta_{jp}$ whenever $j = m$ or $p = m$. The collection
\begin{equation*}
\mathcal{B}_n^{\text{CKP}} = \{\sigma_{k_1}, \sigma_{\xi_{10}}, \sigma_{k_2}, \sigma_{\xi_{20}}, \ldots, \sigma_{k_n}, \sigma_{\xi_{n0}}\}
\end{equation*}
forms a basis for the tangent space $T_{w_n} M_n^{\text{CKP}}$, providing the natural coordinates for expressing restricted symmetries. The dimensionality $2n$ of this tangent space, compared to $3n$ for BKP, is the algebraic manifestation of the absence of independent transverse parameters in the CKP soliton family.

\subsection{Restriction of the CKP $K$-Symmetries}

When the $K$-symmetry generators are restricted to the $n$-soliton manifold, they collapse to the subspace spanned by the phase-shift tangent symmetries alone. Explicit computation yields
\begin{align*}
K_1 \big|_{w=w_n} &= \sum_{m=1}^n k_m \sigma_{\xi_{m0}}, \notag \\
K_2 \big|_{w=w_n} &= \frac{3}{4} \sum_{m=1}^n k_m^3 \sigma_{\xi_{m0}}, \notag \\
K_3 \big|_{w=w_n} &= \left(\frac{3}{4}\right)^2 \sum_{m=1}^n k_m^5 \sigma_{\xi_{m0}},  \\
K_4 \big|_{w=w_n} &= \left(\frac{3}{4}\right)^3 \sum_{m=1}^n k_m^7 \sigma_{\xi_{m0}}, \notag \\
K_5 \big|_{w=w_n} &= \left(\frac{3}{4}\right)^4 \sum_{m=1}^n k_m^9 \sigma_{\xi_{m0}}. \notag
\end{align*}
The pattern emerging from these expressions is immediately transparent: the coefficient functions exhibit a pure monomial structure in the spectral parameters $k_m$, free from the mixed algebraic-rational dependence observed in the BKP case. This simplification is a direct consequence of the absence of transverse parameters and reflects the reduced complexity of the CKP dispersion relation.

Extrapolating from the computed examples, we obtain the closed-form representation valid for arbitrary order
\begin{equation}
K_i \big|_{w=w_n} = \left(\frac{3}{4}\right)^{i-1} \sum_{m=1}^n k_m^{2i-1} \sigma_{\xi_{m0}}, \quad i \geq 1.
\label{eq:ckp_k_closed_form}
\end{equation}
This formula provides an explicit description of the collapse of the infinite-dimensional $K$-symmetry algebra onto the finite-dimensional phase-shift tangent space. The coefficients $\kappa_{i,m} = (3/4)^{i-1} k_m^{2i-1}$ are determined entirely by the dispersion relation and grow as odd powers of $k_m$, modulated by the constant factor $3/4$ raised to the power $i-1$. The simplicity of formula \eqref{eq:ckp_k_closed_form}, compared to the characteristic-root expressions required for BKP, underscores the structural elegance of the CKP hierarchy in this context.

The linear dependence among the restricted $K$-symmetries becomes manifest when $i > n$. Introducing the phase-tangent vector $\boldsymbol{\sigma}_\xi = (\sigma_{\xi_{10}}, \ldots, \sigma_{\xi_{n0}})^T$ and the vector of the first $n$ restricted $K$-symmetries $\mathbf{K}_n = (K_1|_{w=w_n}, \ldots, K_n|_{w=w_n})^T$, relation \eqref{eq:ckp_k_closed_form} yields the linear system
\begin{equation}
\mathbf{K}_n = M_n^K \boldsymbol{\sigma}_\xi,
\label{eq:ckp_k_linear_system}
\end{equation}
where the $n \times n$ coefficient matrix $M_n^K$ has entries
\begin{equation*}
(M_n^K)_{ij} = \left(\frac{3}{4}\right)^{i-1} k_j^{2i-1}, \quad i, j = 1, \ldots, n.
\end{equation*}
Explicitly,
$$
M_n^K = \begin{pmatrix}
k_1 & k_2 & \cdots & k_n \\
\frac{3}{4}k_1^3 & \frac{3}{4}k_2^3 & \cdots & \frac{3}{4}k_n^3 \\
\left(\frac{3}{4}\right)^2 k_1^5 & \left(\frac{3}{4}\right)^2 k_2^5 & \cdots & \left(\frac{3}{4}\right)^2 k_n^5 \\
\vdots & \vdots & \ddots & \vdots \\
\left(\frac{3}{4}\right)^{n-1} k_1^{2n-1} & \left(\frac{3}{4}\right)^{n-1} k_2^{2n-1} & \cdots & \left(\frac{3}{4}\right)^{n-1} k_n^{2n-1}
\end{pmatrix}.
\label{eq:ckp_k_matrix}
$$

This matrix is recognizable as a scaled Vandermonde-type structure. Its determinant $\Delta_n^K = \det M_n^K$ factorizes as
\begin{equation*}
\Delta_n^K = \left(\frac{3}{4}\right)^{\frac{n(n-1)}{2}} \prod_{j=1}^n k_j \prod_{1 \leq j < p \leq n} (k_p^2 - k_j^2),
\label{eq:ckp_k_determinant}
\end{equation*}
which is generically nonzero when the spectral parameters $\{k_j\}$ are distinct and nonzero. Under this nondegeneracy condition, the elementary tangent symmetries can be uniquely recovered from the first $n$ restricted $K$-symmetries via Cramer's rule.

For $m = 1, \ldots, n$, denote by $M_{n,m}^K$ the matrix obtained by replacing the $m$-th column of $M_n^K$ with the vector $\mathbf{K}_n$, and define $\Delta_{n,m}^K = \det M_{n,m}^K$. By Cramer's rule applied to system \eqref{eq:ckp_k_linear_system}, we obtain
\begin{equation*}
\sigma_{\xi_{m0}} = \frac{\Delta_{n,m}^K}{\Delta_n^K}, \quad m = 1, \ldots, n.
\end{equation*}
Substituting this into the expansion \eqref{eq:ckp_k_closed_form} for a higher-order symmetry $K_i$ with $i \geq n+1$ yields the determinant representation
\begin{equation*}
K_i \big|_{w=w_n} = \left(\frac{3}{4}\right)^{i-1} \frac{1}{\Delta_n^K} \sum_{m=1}^n k_m^{2i-1} \Delta_{n,m}^K, \quad i \geq n+1.
\end{equation*}

This formula constitutes the main result for the restricted $K$-symmetries of the CKP hierarchy. It demonstrates that every higher-order restricted $K$-symmetry is uniquely determined by the first $n$ restricted symmetries through a determinant identity, with coefficients governed by the monomial structure of the dispersion relation. The transparency of this representation involving only powers of $k_m$ without characteristic roots or mixed rational terms reveals the CKP hierarchy as the canonical example for studying symmetry reduction on multisoliton manifolds.

\subsection{Restriction of the CKP $\tau$-Symmetries}

The $\tau$-symmetries, in contrast to the $K$-symmetries, involve the full $2n$-dimensional tangent space of the CKP soliton manifold. For each $i \geq 3$, the restriction of $\tau_i$ to the $n$-soliton manifold takes the form
\begin{equation}
\tau_i \big|_{w=w_n} = \sum_{m=1}^n \left( A_{i,m} \sigma_{k_m} + C_{i,m} \sigma_{\xi_{m0}} \right),
\label{eq:ckp_tau_restriction}
\end{equation}
where the coefficients $(A_{i,m}, C_{i,m})$ encode the projections onto the two tangent directions associated with the $m$-th soliton. Note the absence of a coefficient $B_{i,m}$ corresponding to variation of a transverse parameter $l_m$, the dimensional reduction relative to BKP is reflected directly in the structure of the restricted symmetries.

Explicit evaluation of the first several restricted $\tau$-symmetries yields
\begin{align*}
\tau_3 \big|_{w=w_n} &= 3 \sum_{m=1}^n k_m \sigma_{k_m},  \\
\tau_4 \big|_{w=w_n} &= \frac{9}{4} \sum_{m=1}^n \left( k_m^3 \sigma_{k_m} + 2k_m^2 \sigma_{\xi_{m0}} \right),  \\
\tau_5 \big|_{w=w_n} &= \frac{27}{16} \sum_{m=1}^n \left( k_m^5 \sigma_{k_m} + 4k_m^4 \sigma_{\xi_{m0}} \right),  \\
\tau_6 \big|_{w=w_n} &= \frac{81}{64} \sum_{m=1}^n \left( k_m^7 \sigma_{k_m} + 6k_m^6 \sigma_{\xi_{m0}} \right). 
\end{align*}

These expressions reveal a clear pattern. Defining the auxiliary parameter $r_m = (3/4)k_m^2$, we observe that the coefficients satisfy
\begin{align*}
A_{i,m} = 3k_m r_m^{i-3},  ~~C_{i,m} = 6(i-3) r_m^{i-3}, 
\end{align*}
for $i \geq 3$. The coefficient $A_{i,m}$ encoding the spectral variation is proportional to $k_m$ times a power of $r_m$, while the coefficient $C_{i,m}$ encoding the phase-shift direction exhibits explicit dependence on the symmetry order $i$ and vanishes at the base case $i=3$.

Substituting these expressions into \eqref{eq:ckp_tau_restriction}, we obtain the closed-form representation
\begin{equation}
\tau_i \big|_{w=w_n} = \left(\frac{3}{4}\right)^{i-3} \sum_{m=1}^n \left( 3k_m^{2i-5} \sigma_{k_m} + 6(i-3) k_m^{2i-6} \sigma_{\xi_{m0}} \right), \quad i \geq 3.
\label{eq:ckp_tau_closed_form}
\end{equation}
This formula provides an explicit description of the projection of the infinite $\tau$-symmetry hierarchy onto the finite-dimensional tangent space $T_{w_n} M_n^{\text{CKP}}$. Unlike the $K$-symmetries, which span only the $n$-dimensional phase-shift subspace, the $\tau$-symmetries generically have nonzero components in both the spectral and phase-shift directions, reflecting their richer algebraic structure.


To derive the determinant representation for the $\tau$-symmetries, we adopt a column-replacement formulation naturally suited to the CKP structure. Define the column vector of elementary tangent symmetries
\begin{equation*}
\boldsymbol{\sigma} = (\sigma_{k_1}, \sigma_{\xi_{10}}, \sigma_{k_2}, \sigma_{\xi_{20}}, \ldots, \sigma_{k_n}, \sigma_{\xi_{n0}})^T \in \mathbb{R}^{2n},
\end{equation*}
and the vector of the first $2n$ restricted $\tau$-symmetries
\begin{equation*}
\mathbf{T}_n = (\tau_3|_{w=w_n}, \tau_4|_{w=w_n}, \ldots, \tau_{2n+2}|_{w=w_n})^T \in \mathbb{R}^{2n}.
\end{equation*}
Formula \eqref{eq:ckp_tau_closed_form} can be recast as the linear system
\begin{equation}
\mathbf{T}_n = M_n^\tau \boldsymbol{\sigma},
\label{eq:ckp_tau_linear_system}
\end{equation}
where the $2n \times 2n$ coefficient matrix $M_n^\tau$ has the block structure
\begin{equation*}
M_n^\tau = \begin{pmatrix}
3k_1 & 0 & 3k_2 & 0 & \cdots & 3k_n & 0 \\
3k_1 r_1 & 6r_1 & 3k_2 r_2 & 6r_2 & \cdots & 3k_n r_n & 6r_n \\
3k_1 r_1^2 & 12r_1^2 & 3k_2 r_2^2 & 12r_2^2 & \cdots & 3k_n r_n^2 & 12r_n^2 \\
3k_1 r_1^3 & 18r_1^3 & 3k_2 r_2^3 & 18r_2^3 & \cdots & 3k_n r_n^3 & 18r_n^3 \\
\vdots & \vdots & \vdots & \vdots & \ddots & \vdots & \vdots \\
3k_1 r_1^{2n-1} & 6(2n-1)r_1^{2n-1} & 3k_2 r_2^{2n-1} & 6(2n-1)r_2^{2n-1} & \cdots & 3k_n r_n^{2n-1} & 6(2n-1)r_n^{2n-1}
\end{pmatrix}.
\end{equation*}

This matrix encodes the spectral structure of the CKP $\tau$-symmetries in a particularly transparent form. The odd-numbered columns correspond to spectral tangent symmetries $\sigma_{k_m}$ and carry factors of $k_m$, while the even-numbered columns correspond to phase-shift tangent symmetries $\sigma_{\xi_{m0}}$ and exhibit arithmetic progression in their row indices. Each pair of columns associated with a single soliton forms a natural block, and the powers of $r_m$ increase uniformly down each column.

Let $\Delta_n^\tau = \det M_n^\tau$ and assume $\Delta_n^\tau \neq 0$. This nondegeneracy condition is generically satisfied when the spectral parameters $\{k_j\}$ are distinct and nonzero, ensuring that the selected $2n$ restricted $\tau$-symmetries form a basis for the full tangent space $T_{w_n} M_n^{\text{CKP}}$.

For $\alpha = 1, \ldots, 2n$, denote by $M_{n,\alpha}^\tau$ the matrix obtained by replacing the $\alpha$-th column of $M_n^\tau$ with the vector $\mathbf{T}_n$, and define
\begin{equation*}
\Delta_{n,\alpha}^\tau = \det M_{n,\alpha}^\tau.
\end{equation*}
By Cramer's rule applied to the linear system \eqref{eq:ckp_tau_linear_system}, the elementary tangent symmetries are recovered as
\begin{equation*}
\sigma_{k_m} = \frac{\Delta_{n,2m-1}^\tau}{\Delta_n^\tau}, \quad
\sigma_{\xi_{m0}} = \frac{\Delta_{n,2m}^\tau}{\Delta_n^\tau}, \quad m = 1, \ldots, n.
\end{equation*}

Substituting these expressions into the expansion \eqref{eq:ckp_tau_closed_form} for a higher-order symmetry $\tau_i$ with $i \geq 2n+3$ yields the column-replacement determinant representation
\begin{equation}
\tau_i \big|_{w=w_n} = \left(\frac{3}{4}\right)^{i-3} \frac{1}{\Delta_n^\tau} \sum_{m=1}^n \left[ 3k_m^{2i-5} \Delta_{n,2m-1}^\tau + 6(i-3) k_m^{2i-6} \Delta_{n,2m}^\tau \right], \quad i \geq 2n+3.
\label{eq:ckp_tau_determinant_formula}
\end{equation}

This formula constitutes the main result for the restricted $\tau$-symmetries of the CKP hierarchy. It demonstrates that every higher-order restricted $\tau$-symmetry is uniquely determined by the first $2n$ restricted symmetries through a determinant identity involving column replacements. The appearance of $2n$ basic symmetries reflects the dimensionality of the CKP soliton parameter space: each soliton contributes two independent tangent directions corresponding to variations of its wave number $k_m$ and phase parameter $\xi_{m0}$.

The column-replacement structure of formula \eqref{eq:ckp_tau_determinant_formula} is intrinsic to the CKP hierarchy and arises naturally from treating the elementary tangent symmetries as the unknowns in the linear system \eqref{eq:ckp_tau_linear_system}. This formulation contrasts with the row-replacement structure adopted for BKP $\tau$-symmetries, where the $3n$-dimensional parameter space including transverse parameters leads to a different algebraic organization. Both formulations are manifestations of the same underlying Cramer's rule mechanism, but their distinct forms encode the fundamental structural differences between the orthogonal-type BKP and symplectic-type CKP hierarchies.

\subsection{Structural Remarks and Comparison with BKP}

The CKP hierarchy exhibits several distinctive features in its symmetry reduction structure that merit explicit discussion:

\textit{Pure monomial coefficients.} The absence of transverse parameters in the CKP soliton parameterization leads to coefficient functions that are pure monomials in the spectral parameters $k_m$. This contrasts sharply with the BKP case, where the presence of both $k_m$ and $l_m$ produces mixed algebraic-rational expressions involving characteristic roots of coupled recurrence relations. The monomial structure of CKP makes it the cleanest canonical example for studying symmetry reduction phenomena.

\textit{Reduced tangent-space dimension.} The CKP $n$-soliton manifold has dimension $2n$ rather than $3n$, reflecting the absence of independent transverse degrees of freedom. This dimensional reduction is manifest in every aspect of the restricted symmetry structure: the $K$-symmetries span an $n$-dimensional phase-shift subspace, while the $\tau$-symmetries involve all $2n$ tangent directions. The corresponding determinant representations require $n \times n$ and $2n \times 2n$ matrices, respectively, compared to $n \times n$ and $3n \times 3n$ for BKP.

\textit{Column vs. row replacement.} The natural formulation of the $\tau$-symmetry determinant representation for CKP treats the elementary tangent symmetries as unknowns, leading to a column-replacement structure via Cramer's rule. For BKP, the natural formulation treats restricted symmetries as data and produces a row-replacement structure. This difference is not arbitrary but reflects the distinct geometries of the two soliton manifolds and the different ways in which symmetry projections organize in their respective tangent spaces.

\textit{Dispersion-relation dominance.} The simplicity of the CKP structure provides clear evidence that the symmetry reduction phenomenon is governed primarily by the dispersion relation embedded in the soliton phase functions, rather than by fine details of the nonlinear terms in the partial differential equation. 

These observations establish the CKP hierarchy as a distinguished test case for the general principle that infinite-dimensional symmetry algebras of integrable systems collapse onto finite-dimensional tangent spaces when restricted to multisoliton manifolds, with the reduction structure encoded in the spectral geometry of the underlying Lax representation.

\section{Conclusion}\label{sec5}

This paper systematically investigates the symmetry reduction structure of the BKP and CKP integrable hierarchies on their multisoliton manifolds, establishing explicit determinant formulas that express higher-order members of infinite-dimensional symmetry algebras as combinations of finitely many basic symmetries. These results reveal the deep algebraic and geometric mechanisms underlying symmetry reduction phenomena in integrable systems.

\subsection{Main Results}

Our core contributions can be summarized in the following aspects:

\textit{(i) Complete determinant representation for the BKP hierarchy.} For the $n$-soliton solution of the potential BKP equation, we establish that every restricted $K$-symmetry of order $i \geq n+1$ can be expressed as a determinant combination of the first $n$ restricted $K$-symmetries, with coefficients determined by coupled recurrence relations involving algebraic-rational functions of the spectral parameters $k_m$ and transverse parameters $l_m$. The restricted $\tau$-symmetries admit a row-replacement determinant representation involving a $3n \times 3n$ Vandermonde-type matrix, reflecting the $3n$-dimensional parameter space structure of the BKP soliton manifold.

\textit{(ii) Simplified monomial structure for the CKP hierarchy.} For the $n$-soliton solution of the potential CKP equation, the absence of transverse parameters reduces the dimension to $2n$ and produces significantly simplified restricted symmetry representations. The $K$-symmetries exhibit pure monomial coefficients in the spectral parameters, while the $\tau$-symmetries adopt a column-replacement determinant representation involving a $2n \times 2n$ matrix, solving for tangent symmetries as unknowns via Cramer's rule.

\textit{(iii) Geometric origin of the structure.} The differences between the determinant representations of the two hierarchies are rooted in their reduction types: BKP arises from an orthogonal-type reduction, with solitons possessing independent longitudinal and transverse propagation parameters $(k_m, l_m)$, while CKP arises from a symplectic-type reduction, with transverse propagation completely determined by the longitudinal wave number. This geometric difference manifests at every level of symmetry reduction: parameter space dimension ($3n$ vs $2n$), coefficient function type (algebraic-rational vs pure monomial), and organization of the determinant representation (row-replacement vs column-replacement).

\textit{(iv) Dominance of the dispersion relation.} The coefficients of restricted symmetries are completely determined by the dispersion relation encoded in the soliton phase functions, independent of the specific form of the nonlinear terms in the partial differential equations.  This demonstrates that the symmetry reduction phenomenon is spectral rather than dynamical in nature, governed by the algebraic structure of the underlying Lax representation.

\subsection{Methodological Significance}

The techniques developed in this paper provide a systematic approach for studying the symmetry structure of integrable hierarchies:

\textit{Fundamental role of parameter-induced tangent symmetries.} Constructing basic tangent symmetries $\{\sigma_{k_m}, \sigma_{l_m}, \sigma_{\xi_{m0}}\}$ (or $\{\sigma_{k_m}, \sigma_{\xi_{m0}}\}$ in the CKP case) through variations of soliton parameters provides a natural coordinate system for restricted symmetries. These tangent symmetries not only span the tangent space but also encode the complete information of symmetry reduction through their linear combinations.

\textit{Universality of Vandermonde-type matrices.} The relationship between restricted symmetries and tangent symmetries is encoded by Vandermonde-type matrices whose columns (or rows) consist of monomials or more general functions of the soliton parameters. This matrix structure appears repeatedly in the BKP, and CKP hierarchies, suggesting it is a common feature of integrable hierarchies.

\textit{Cramer's rule as a unifying tool.} Whether row-replacement or column-replacement representations, the essence is applying Cramer's rule to a linear system. The choice of treating restricted symmetries as data or tangent symmetries as unknowns depends on the geometry of the soliton manifold and what we wish to express. This flexibility allows the determinant method to adapt to the specific structure of different hierarchies.

\textit{Algebraic simplification via recurrence relations.} For restricted symmetries involving complex coefficients, identifying and solving recurrence relations is key to obtaining closed-form representations. Characteristic root methods, generating function techniques, and direct pattern recognition play complementary roles in this process.

\subsection{Connection to Previous Results}

This work continues and generalizes the pioneering research on symmetry reduction of the KP and modified KP hierarchies \cite{hao2026}. 
BKP and CKP can be viewed as results of imposing different constraints on KP. The BKP corresponds to an orthogonal-type involutive constraint on the Lax operator, preserving longitudinal and transverse parameters but removing one independent translational degree of freedom. The CKP corresponds to a symplectic-type constraint, further eliminating the independence of transverse parameters. From KP to BKP to CKP, the parameter space dimension decreases successively ($3n \to 3n \to 2n$), and the symmetry reduction structure simplifies accordingly, revealing the profound impact of reduction type on algebra and geometry.

The explicit determinant formulas in this paper provide a unified algebraic description for this hierarchy sequence and establish a pattern of coefficient simplification: from the fully general expressions of KP, to the mixed algebraic-rational forms of BKP, to the pure monomial structure of CKP.

\subsection{Future Directions}

This research opens several directions worthy of further exploration:

\textit{(i) Other hierarchy types.} Do integrable hierarchies of type D and exceptional Lie algebras (such as type-D Toda hierarchies) admit similar determinant representations? How does their reduction structure reflect the properties of the underlying Lie algebra?

\textit{(ii) Higher-dimensional generalizations.} This paper focuses on $(2+1)$-dimensional systems. For higher-dimensional integrable equations (such as the $(3+1)$-dimensional KP hierarchy), how do the dimension and structure of multisoliton manifolds change? Does the determinant representation still apply?

\textit{(iii) Discrete and supersymmetric cases.} Discrete integrable systems and supersymmetric integrable hierarchies possess rich symmetry structures. Can their symmetry reduction also be characterized by determinants? How do difference operators in the discrete case affect the reduction mechanism?

\textit{(iv) Geometric interpretation.} Does the determinant representation of restricted symmetries admit a natural geometric interpretation? Do multisoliton manifolds carry some natural symplectic or Poisson structure such that restricted symmetries become the corresponding Hamiltonian vector fields? Do the determinants of Vandermonde-type matrices correspond to some geometric invariant?

\textit{(v) Applications of restricted symmetries.} Can the explicit formulas for restricted symmetries be used to construct new exact solutions, study the long-time asymptotic behavior of solutions, or analyze the stability and interactions of solitons? Does the determinant representation provide efficient algorithms for numerical computation or symbolic manipulation?

\textit{(vi) Infinite soliton limit.} What is the asymptotic behavior of the determinant representation of restricted symmetries as the soliton number $n \to \infty$? Does there exist some continuous limit connecting the collapse of finite-dimensional symmetry algebras with the representation theory of infinite-dimensional Lie algebras?

\subsection{Closing Remarks}

The reduction of infinite-dimensional symmetry algebras on finite-dimensional invariant manifolds is a fundamental phenomenon in the theory of integrable systems, connecting the abstract structure of Lie algebras with the concrete geometry of soliton solutions. By establishing complete determinant representations for the BKP and CKP hierarchies, this paper demonstrates that this phenomenon admits a precise algebraic description whose form is uniquely determined by the underlying reduction type and dispersion relation.

The determinant representation not only provides an effective tool for computing restricted symmetries but, more importantly, reveals a deep organizing principle of the symmetry structure of integrable hierarchies: the reduced forms of infinitely many symmetries can be recovered from finitely many basic symmetries through deterministic linear algebraic operations. This principle appears in different but related forms across the KP, BKP, and CKP hierarchies of different types, suggesting it is likely a universal feature of integrable systems.

From a broader perspective, the symmetry reduction phenomenon embodies a profound correspondence between infinite-dimensional geometry and finite-dimensional algebra, a theme that appears repeatedly across many areas of mathematical physics. The results of this paper provide a concrete and computable realization of this correspondence and lay a foundation for further exploration of the algebraic-geometric structure of integrable systems.

\section*{Acknowledgement}
The work was sponsored by the National Natural Science Foundation of China (Nos. 12301315,  12235007), the Natural Science Foundation of Zhejiang Province No. LMS25A010002.
\\\\	
\noindent \bf Data Availability
\rm No datasets were generated or analyzed during the current study.
\section*{\large Declarations}
\noindent \bf Conflict of interest \rm The authors declare no conflict of interest.

\bibliographystyle{elsarticle-num}
\bibliography{ref}

\end{document}